\documentclass{ieeeaccess}
\usepackage{cite}
\usepackage{amsmath,amssymb,amsfonts}
\usepackage{algorithmic}
\usepackage{graphicx}  % Only load this once
\usepackage{textcomp}
\definecolor{blue2}{rgb}{0,0,0.6078}
\definecolor{Red}{rgb}{1, 0, 0}
\definecolor{Green}{rgb}{0, 0.5, 0}
\definecolor{blue}{rgb}{0, 0, 1}  % lowercase blue also needs defining sometimes

\usepackage{subcaption}

\def\BibTeX{{\rm B\kern-.05em{\sc i\kern-.025em b}\kern-.08em
    T\kern-.1667em\lower.7ex\hbox{E}\kern-.125emX}}
\begin{document}

\history{This work has been submitted to IEEE for possible publication. Copyright may be transferred without notice, after which this version may no longer be accessible. Date of publication xxxx 00, 0000, date of current version xxxx 00, 0000. \doi{https://arxiv.org/pdf/2411.11382}}

% \history{Date of publication xxxx 00, 0000, date of current version xxxx 00, 0000.}
% \doi{10.1109/ACCESS.2025.DOI}

% \pagenumbering{Arabic} 
\title{Quantifying Haptic Affection of Car Door through Data-Driven Analysis of Force Profile}

\author{\uppercase{Mudassir Ibrahim Awan}\authorrefmark{1*},
\uppercase{Ahsan Raza\authorrefmark{1*}, Waseem Hassan\authorrefmark{1,2}, Ki-Uk Kyung\authorrefmark{3} and Seokhee Jeon}.\authorrefmark{1}
}
\address[1]{Department of Computer Engineering, Kyung Hee University (e-mail: [miawan, ahsanraza, jeon]@khu.ac.kr)}
\address[2]{Department of Computer Science, University of Copenhagen, Denmark (e-mail: waha@di.ku.dk)}
\address[3]{Department of Mechanical Engineering, the Korea Advanced Institute of Science and Technology, Daejeon 34141, South Korea (e-mail: kyungku@kaist.ac.kr)}

\address[*]{These authors contributed equally to this paper.}

% \tfootnote{The acknowledgment goes here}

\markboth
{Author \headeretal: Preparation of Papers for IEEE Access}
{Author \headeretal: Preparation of Papers for IEEE Access}

% \corresp{Corresponding author: First A. Author (e-mail: author@ boulder.nist.gov).}
\corresp{Corresponding author: Seokhee Jeon (e-mail: jeon@khu.ac.kr).}

\begin{abstract}
Haptic affection plays a crucial role in user experience, particularly in the automotive industry where the tactile quality of components can influence customer satisfaction. This study aims to accurately predict the affective property of a car door by only watching the force or torque profile of it when opening. To this end, a deep learning model is designed to capture the underlying relationships between force profiles and user-defined adjective ratings, providing insights into the door-opening experience. The dataset employed in this research includes force profiles and user adjective ratings collected from six distinct car models, reflecting a diverse set of door-opening characteristics and tactile feedback. The model's performance is assessed using Leave-One-Out Cross-Validation, a method that measures its generalization capability on unseen data. The results demonstrate that the proposed model achieves a high level of prediction accuracy, indicating its potential in various applications related to haptic affection and design optimization in the automotive industry.
\end{abstract}


%\titlepgskip=-15pt

\maketitle

\section{Introduction}
\label{sec:introduction}

The automotive industry increasingly emphasizes user experience, focusing on the physical sensations and emotions drivers and passengers feel when interacting with different aspects of a vehicle \cite{norman2004emotional, hekkert2006design}. 
Among these interactions, the tactile experience of operating a car door is crucial, as it serves as the first point of contact between the user and the vehicle \cite{kim2018sound}. 
The way a door feels when opened or closed can leave a lasting impression on the overall perception of the car's quality and craftsmanship.

Given the importance of this initial interaction, car designers would greatly benefit from a virtual evaluation system capable of predicting the affective response users may have when interacting with car doors. 
Predicting user perceptions based on early-stage design information \textemdash such as the physical properties of door components like hinge profiles and force/torque distributions \textemdash could streamline the design process by reducing the reliance on physical prototypes.

\begin{figure*}[t]
        \centering
         \includegraphics[width = 2\columnwidth]{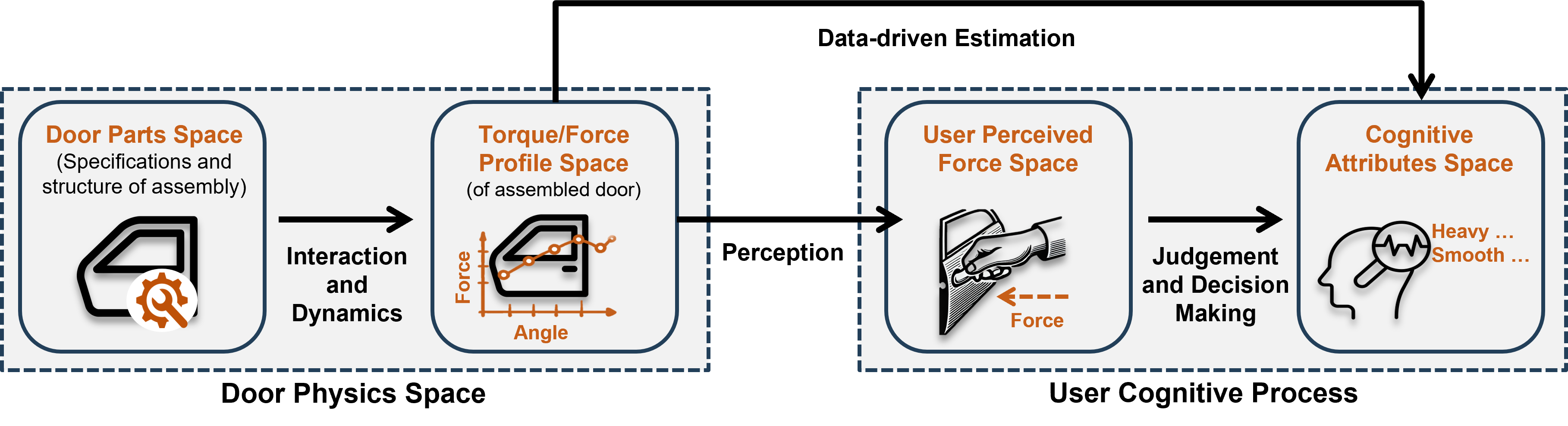}
        \caption{Schematic representation of the sequential pipeline linking physical attributes of a car door to user perceptual experience.}
        \label{fig:pipeline}
\end{figure*}

In automotive design, the physical attributes of a car door significantly influence user experience \cite{bezat2007car, strolz2010simulation}. 
Similar research in haptic feedback devices and consumer products has shown that the tactile qualities of interfaces, such as knobs and buttons, directly affect user satisfaction and perceived quality \cite{van2024substitute}. 
However, the relationship between the physical attributes of a car door and their effects on affection remains elusive, requiring a more in-depth analysis \cite{leder2005dimensions}.
Moreover, the tactile sensation of the vehicle interface can significantly vary depending on factors such as assembly parts, tolerances between components, and wear conditions \cite{parizet2008analysis}.
Developing a system that connects these physical attributes to the user’s perceptual experience can be highly valuable. 
This relationship can be conceptualized as a sequential pipeline (see Fig. \ref{fig:pipeline}), which begins with the specifications of car door components and leads to the final user experience.

The first stage, \textit{Door Parts Space}, represents the physical specifications and structural attributes of the door components (e.g., total weight, center of mass, hinge profile, joint friction, etc.). 
These attributes contribute to the dynamic behavior of the door during its operation, such as opening and closing, which can be described as a \textit{Torque and Force Profile}. 
The physical dynamics of the door determine this profile \cite{strolz2010simulation, strolz2008haptic}.
The relationship between the components of the car door and its profile is well-studied and remains common knowledge in the automotive industry \cite{strolz2009control, raghuveer2014design}. 

The torque profile is then converted into a \textit{User Perceived Force}, representing the force felt by the user while operating the door. 
This conversion can be simulated or recorded through sensors \cite{strolz2009design, ma2024hybrid}. 
Mechanoreceptors in the skin and joints mediate this force perception, providing a predictable link between the physical door and user interaction \cite{deflorio2022skin}.

However, the subsequent step—converting the user’s perceived force into \textit{Cognitive Attributes} (e.g., judgments of comfort, smoothness, or quality)—involves the user's cognitive process, making it less predictable. 
Human cognitive perception is influenced by individual differences, prior experiences, and contextual factors, leading to variability in subjective evaluations \cite{aru2016early}.
Therefore, this step requires well-designed perception studies involving human participants to model these subjective evaluations accurately.

Within this pipeline, much of the existing research focuses on the relationship between door components and torque/force profiles \cite{gheorghe2021new} \textemdash the \textit{Door Physics Space}.
Although some studies have analyzed cognitive characteristics \cite{chang2016kansei}, there is limited research on how these \textit{Door Parts Space} translate into the \textit{User Cognitive Processes}.
In other words, while the physical aspects of the door's operation are well understood, the crucial step of predicting user experience remains elusive. 
Successfully estimating the conversion from the \textit{Door Physics Space} to the \textit{Cognitive Attribute Space} would make such predictions possible.

While it is theoretically possible to establish a mapping from the \textit{Door Parts} to the \textit{Cognitive Attributes}, it is impractical due to the large number of permutations required (every single door part has to be mapped). 
A more practical approach is to focus on the relationship between \textit{Torque/Force Profile Space} and the \textit{Cognitive Attributes Space}. 
As indicated by the \textit{Data-driven Estimation} step in the pipeline (Fig. \ref{fig:pipeline}), using real force profile data recorded from car doors allows us to apply data-driven methods to estimate user cognition more effectively. 
This approach simplifies the mapping process by aggregating the effects of various door components into a single force profile.

To address the challenges and bottlenecks in the field, researchers have turned to machine learning as a potential solution. 
Advanced algorithms can identify patterns and relationships that may be difficult for humans to discern \cite{khanal2021classification, ferreira2022conformity, lai2022kansei}. 
Machine learning models, particularly deep learning architectures, have shown promise in modeling complex, non-linear relationships between input data and user perceptions \cite{wang2022data, hassan2023establishing, awan2023design}. 
However, there has been limited research on leveraging these techniques to predict human perception of car door attributes based on vehicle data.

In this study, we propose a machine learning approach to predict users' haptic perception from car door force profiles. 
We use a CNN-LSTM network, where the CNN extracts spatial patterns and the LSTM captures temporal dependencies in the force profile data. 
Participants rated their experiences using antonymously paired haptic adjectives (such as, ``smooth–rough'', ``heavy–light'') commonly used in haptic perception research \cite{Gao2016Deep, richardson2020learning, hassan2023establishing}.
The trained network was validated through cross-validation, demonstrating its effectiveness in predicting user cognitive attributes based on force profile data. 
Our system can be integrated into automotive design workflows, providing early predictive insights to optimize designs and improve user experience while reducing costs.

\section{Related Works}
\label{sec:relatedworks}
This section provides an overview of the existing literature related to the perception of car doors and the use of machine learning in the automotive industry. 

\subsection{Machine Learning and Perception of Cars/Car Parts}
Machine learning techniques are used in various aspects of automotive design, such as comfort, aesthetics, and usability \cite{khanal2021classification, lai2022kansei}.
By training models on large datasets containing information about car designs and user feedback, these studies have been able to identify patterns and relationships that can inform the design process.

Machine learning has shown particular promise in the prediction of users' emotional responses to car designs. 
Researchers have developed models that can accurately predict users' emotional reactions to different car designs based on features such as color and shape \cite{lai2022kansei}. 
This has provided insights into the emotional aspects of automotive design and has the potential to inform the creation of more emotionally engaging vehicles.
Machine learning has been used to analyze the relationship between the physical properties of cars and their perceived quality, such as the perceived quality of sound produced by the engine \cite{duvigneau2016analysis, muender2022howl}. 
However, the application of machine learning in the context of predicting haptic perception and emotions related to car doors remains a relatively unexplored area.

\begin{figure*} 
        \centering
         \includegraphics[width = 2\columnwidth, height = 8cm]{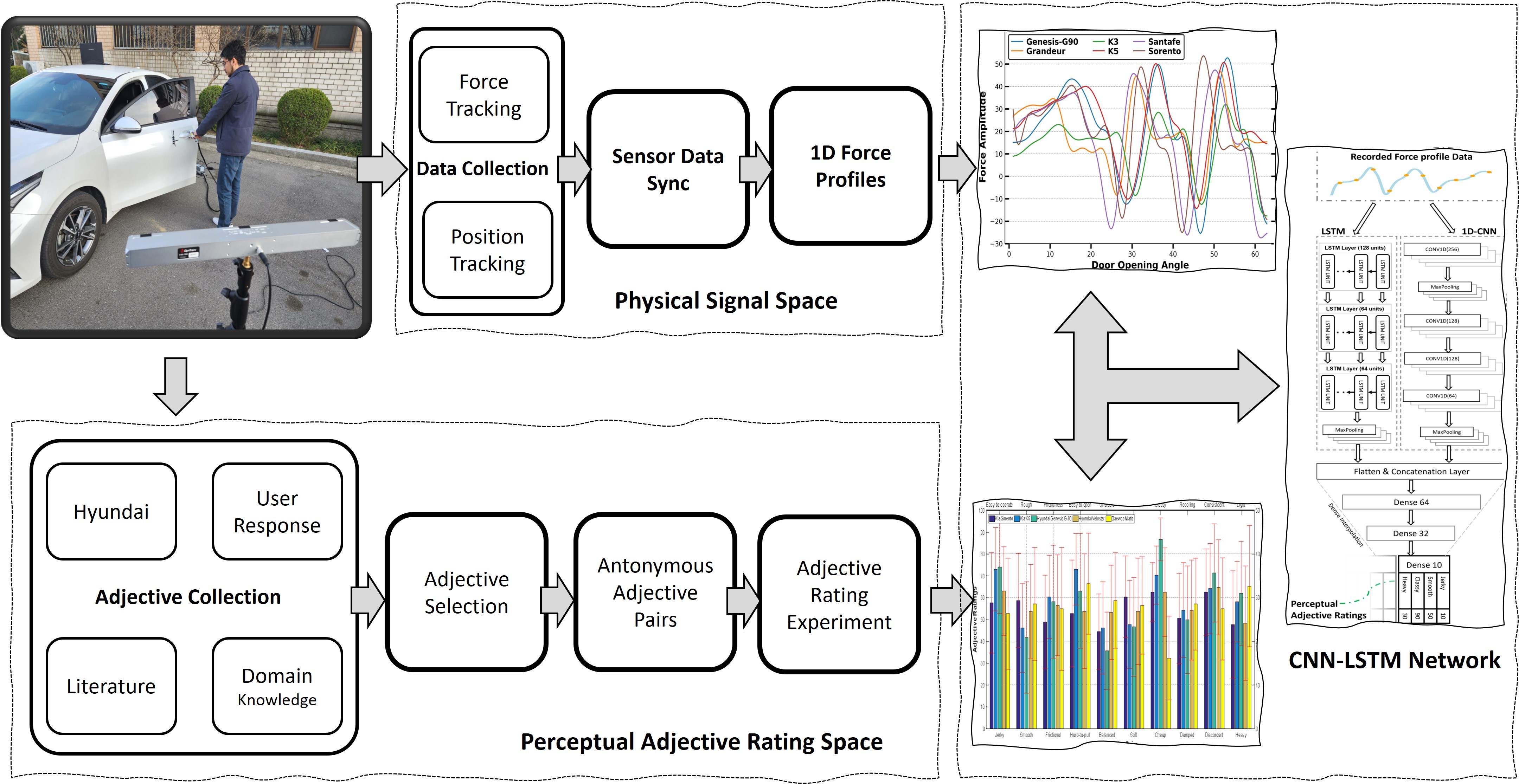}
        \caption{An overview of the overall study. Experiments with a car door provide the force and position tracking values as well as user ratings for the perception of opening a car door. These data are used to train a CNN-LSTM model that predicts perceived ratings based on force profiles of opening a car door.}
        \label{fig:framework}
\end{figure*}

\subsection{The Role of Emotions in Product Design}
Emotions play a crucial role in shaping users' perception of products and their overall satisfaction \cite{elkharraz2013making, desmet2018measuring}. 
Affective engineering has emerged as an interdisciplinary field that aims to incorporate users' emotions and preferences into the design process, thereby enhancing the overall user experience \cite{wu2022pleasurable}. 
Some studies have explored the role of emotions in the context of automotive design, focusing on various aspects such as the interior environment, the driving experience, and the vehicle's appearance \cite{cha2019affective}. 
However, there is still limited research on the role of emotions in the design of car doors and their impact on users' haptic perception and satisfaction.

Car door design plays a critical role in the overall user experience of a vehicle. 
Early research in this area focused on the optimization of car door dynamics, with an emphasis on improving the opening and closing characteristics \cite{grujicic2009multi}. 
This body of work has led to the development of various techniques and approaches for optimizing car door design, such as the use of advanced materials and manufacturing processes.

Recently, there has been a shift in focus toward understanding the relationship between car door design and perception.
Studies have explored the impact of car door design on users' perception of quality and luxury \cite{wang2022data}. 
These investigations have revealed that users associate certain design elements, such as the smoothness of the door opening and closing motion, with higher-quality vehicles.
By adopting such an approach, designers can create car doors that not only perform well in terms of functionality but also evoke positive emotions and contribute to an overall satisfying user experience.

\subsection{Haptic Perception in Automotive Design}
Haptic perception, the sense of touch, plays a critical role in how users experience and interact with products \cite{lederman2009haptic, carbon2012model}. 
In automotive design, haptic perception encompasses not only the tactile sensations experienced when touching surfaces and materials \cite{ajovalasit2019human} but also the kinesthetic feedback associated with operating mechanisms \cite{yin2013haptic, katzourakis2014haptic}. 
A better understanding of haptic perception can help designers create more satisfying and user-friendly experiences \cite{lindemann2019methodical}. 
Despite its importance, research on haptic perception in automotive design has been limited, with few studies exploring the factors that contribute to the perception of car door quality and the emotions they evoke.
By using machine learning techniques, designers can create more intuitive and engaging interfaces that cater to the diverse preferences and needs of users.

\subsection{Data-driven Approaches in Automotive Design}
Data-driven approaches have gained traction in various fields, including automotive design, where they enable designers to make informed decisions based on empirical data \cite{ebel2020role, ebel2021automotive, fridman2019advanced}. 
Researchers have used data-driven models for various aspects of automobiles, such as improving the braking control systems \cite{formentin2011data}, or evaluating the health of electronic systems on board \cite{sankavaram2009model}.

Data-driven methods, combined with machine learning techniques, can facilitate the development of predictive models that account for the complex relationships between product properties \cite{du2018new}. 
Despite the potential benefits, there is still a need for more research on data-driven approaches in the context of automotive design, especially regarding car door perception and the emotions they trigger.

\section{Overview}
\label{sec:overview}

Figure \ref{fig:framework} provides a summary of the different sections detailed in this paper.
This section presents a concise version of the paper's content, outlining the main
topics in each of the sections. 
Details are provided in the subsequent sections.

The main aim of the current study is to provide a method to designers and engineers for predicting the cognitive attributes of car doors without the need for prototyping.
We use the terms perception and perceptual attributes interchangeably with cognitive attributes throughout the text.
To this end, Sec. \ref{sec:perceptualadjectiveratingexperiment} details the dataset, experiments, and their procedures for quantifying the perception of opening a car door. 
The experiments start with collecting a diverse corpus of adjectives to describe the perception of opening a car door, proceed with selecting a limited and more relevant set, and conclude with user ratings across a set of antonymously paired adjectives. 

The force and optical data collection setup for generating the force profiles of opening car doors is explored in Sec. \ref{sec:forceprofile}.
The force profiles portray the amount of force required at various stages of opening a door and are therefore represented as a function of force and the angle of opening.

The data generated from user ratings and force profiles are used as input to train a CNN-LSTM network, presented in Sec. \ref{sec:cnnlstmnetwork}.
The trained CNN-LSTM model can predict the haptic perception of car doors based on their force profiles.
The predicting ability of the network is tested using LOOCV (leave-one-out cross-validation) in Sec. \ref{sec:evaluation}.

\section{Perceptual Adjective Rating Experiment}
\label{sec:perceptualadjectiveratingexperiment}
The aim of this experiment was to describe the act of opening a car door from a perceptual experience point of view.
Users provided ratings against a set of attributes that describe the perception of opening a car door.
The overall experiment can be divided into three sub-experiments which were conducted sequentially.
First, users were asked to open a car door and provide adjectives that can describe the perception of opening a car door.
These adjectives, along with adjectives gathered from literature and other sources, were pooled together to form a lexicon of adjectives.
In the second experiment, users selected the most appropriate adjectives from the lexicon of adjectives.
In the third experiment, users rated the act of opening a car door against the selected adjectives in experiment two.
Details of the dataset and all experiments are provided in the following subsections.

\subsection{Participants and Dataset}
\label{subsec:participantsanddataset}
A total of 20 participants took part in the first and second experiments, and 26 in the third.
Around 75$\%$ of the participants in all experiments were common, the remaining were replaced due to non-availability. 
The majority of the participants identified as males, while 10 out of the combined 66 across all experiments identified as females. 
Their average age was 27.5 years (range: 21 - 34).
None of the participants reported any disabilities or any other factors that could prevent them from successfully participating in the experiments. 
All participants were compensated with $\$$15 USD per experiment.

A total of six cars were used in this experiment.
A wide variation of cars was included in the dataset to cover the range from luxury to utility cars.
The six cars used in the experiment were the new K3 (Kia), K5 DL3 (Kia), the new Grandeur (Hyundai), Genesis G90 (Hyundai), Santafe 7-seater (Hyundai), and Sorento 5-seater (Kia).

\begin{table}
\centering
\caption{The lexicon of adjectives built from four sources, i.e., \textcolor{Green}{Hyundai research}, Experiment, \textcolor{red}{literature}, and \textcolor{blue}{domain expert}. 
The overall list was formed as a result of experiments 1 and 2.}
\label{tab:lexiconofadjectives}
\begin{tabular}{|l|l|l|l|}
\hline
1   \textcolor{Red}{Agitating}    & 18  \textcolor{Green}{\begin{tabular}[c]{@{}l@{}}Easy to \\ operate\end{tabular}}   & 35  \textcolor{red}{Harmonic}  & 52  \textcolor{Green}{\begin{tabular}[c]{@{}l@{}}Cheerful, \\ rhythmical\end{tabular} }\\
2   \textcolor{blue}{Archaic}      & 19  Effortless      & 36  Heavy     & 53  Rigid                                                           \\
3   Balanced     & 20  \textcolor{red}{Empty}           & 37  High      & 54  Rough                                                           \\
4  Calm         & 21  \textcolor{blue}{Erratic}         & 38  Jarring   & 55  Shaking                                                         \\
5  \textcolor{red}{ Calming}      & 22  \textcolor{red}{Exciting}        & 39  Jerky     & 56  \textcolor{blue}{Shallow}                                                         \\
6   Cheap        & 23  Expensive       & 40  \textcolor{red}{Joyful}    & 57  Smooth                                                          \\
7  \textcolor{blue}{ Classy}       & 24  \textcolor{Green}{Stepwise}        & 41  Light     & 58  Soft                                                            \\
8   \textcolor{Green}{Clinging}     & 25 \textcolor{blue}{Fluctuating}     & 42  \textcolor{Green}{Like new}  & 59  \textcolor{Green}{Sophisticated}                                                   \\
9   Comfortable  & 26  \textcolor{blue}{Fluid  }         & 43  \textcolor{blue}{Loud}      & 60  Stiff                                                           \\
10  \textcolor{blue}{Consistent}   & 27  Forceful        & 44  Luxurious & 61 \textcolor{red}{Stressing}                                                       \\
11  \textcolor{Green}{Constant}     & 28 \textcolor{blue}{ Free}            & 45  \textcolor{Green}{Natural}   & 62  Stuck                                                           \\
12  Cool         & 29  \textcolor{blue}{Frictional}      & 46  Not fit   & 63  Tightly fit                                                     \\
13  Damped       & 30  Frictionless    & 47  Old       & 64  Uncomfortable                                                   \\
14  \textcolor{red}{Discordant}   & 31  \textcolor{blue}{Futuristic}      & 48  \textcolor{red}{Pleasant}  & 65  Unpleasant                                                      \\
15  \textcolor{red}{Disturbing}   & 32  \textcolor{red}{Gloomy}          & 49  Quiet     & 66  \textcolor{blue}{Unstable   }                                                     \\
16  \textcolor{blue}{Easy}         & 33  Hard            & 50  Recoiling & 67  \textcolor{red}{Vibrating}                                                       \\
17  \begin{tabular}[c]{@{}l@{}}Easy to \\ open\end{tabular} & 34  \begin{tabular}[c]{@{}l@{}}Hard to \\ pull\end{tabular}  & 51  \textcolor{red}{Relaxing}  & 68  \textcolor{blue}{Vintage }                                                        \\ \hline
\end{tabular}
\end{table}

\subsection{Experiment 1: Lexicon of Adjectives}
\label{subsec:experiment1}
The aim of this experiment was to gather all possible adjectives that can be used to describe the perception of opening a car door. 
A total of four different sources were used to establish the lexicon of adjectives.
These sources were literature, the research provided by Hyundai, the authors' intuition/domain knowledge, and a user experiment.

In the experiment, users were asked to open the driver-side front door with their left hand and open it all the way. 
They were allowed to repeat the procedure if needed, and there were no time restraints. 
The users were handed a paper to write down all the adjectives that could describe the perception of opening a door.
They were informed that they could comment on the motion of the door, its perceptual aesthetics, overall feel, or any other aspect they deemed important.
Every user repeated this process for all the cars.

\begin{figure}[t]
        \centering
        \includegraphics[width=1\columnwidth]{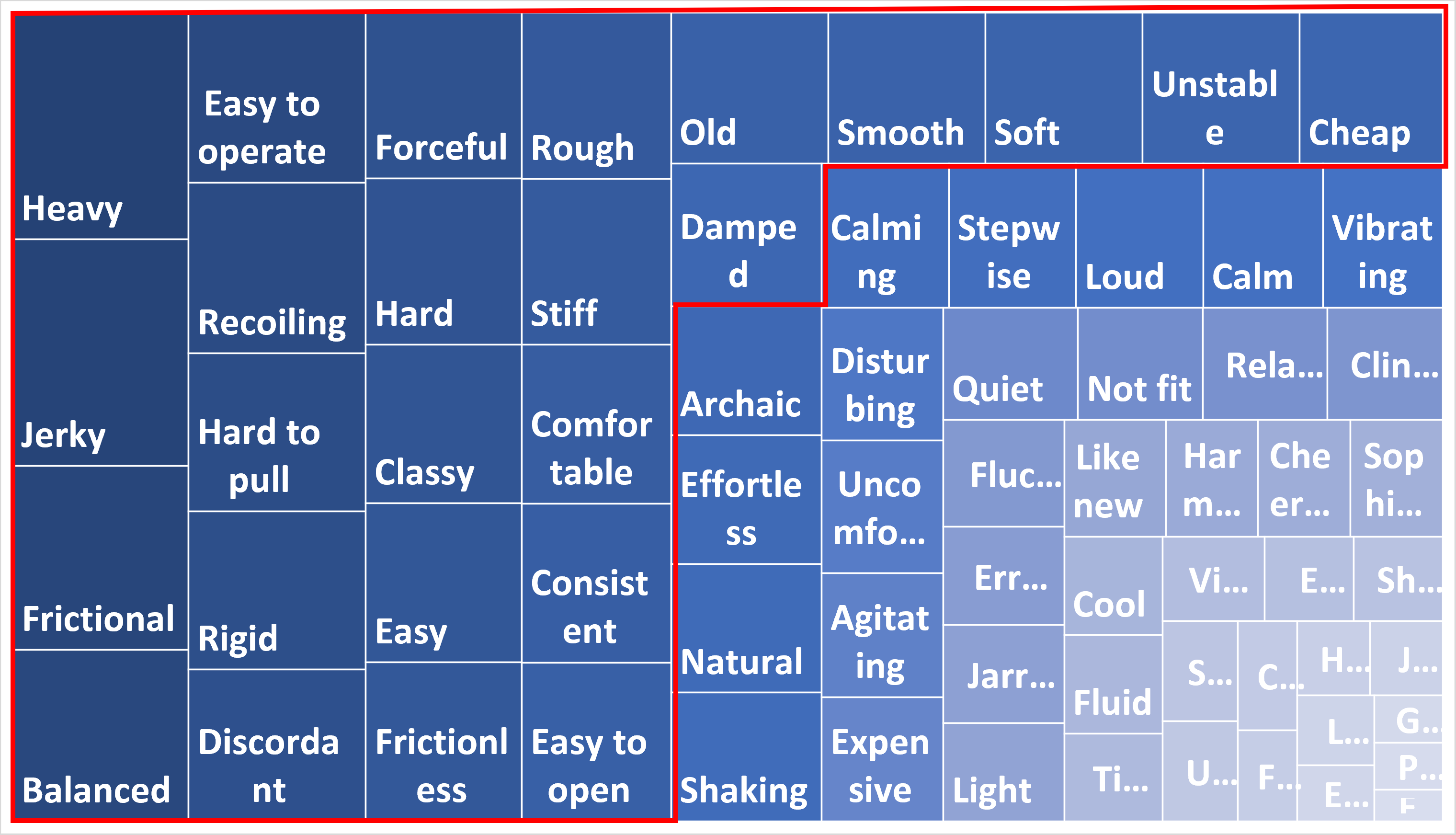}
        \caption{Relevance of all adjectives shown in percentage. The sizes of the boxes are sorted according to relevance percentage and the red border outlines the adjectives that were considered as relevant by at least 20$\%$ of the users.}
        \label{fig:relevantadjectives}
\end{figure}

\subsection{Experiment 2: Selection of Adjectives}
\label{subsec:experiment2}

The main aim of this experiment was to select the most relevant adjectives that describe the perception of opening a car door. 
The lexicon of adjectives contained 68 adjectives, and it was not feasible or productive to continue with all 68.
This experiment was conducted to filter out the adjectives that users considered relevant. 

The users were asked to engage with the door of a car and open it at will. 
They were provided with a list of all the adjectives collected after the first two experiments.
The users had to decide whether a particular adjective was relevant to opening the door of a specific car.
The decision was either a 1 for yes or a 0 for no.
All the users provided their own list of relevant adjectives for each car.

% \textbf{Results of Experiment 1 and 2:}

\vspace{1em}
\noindent \textbf{Results of Experiment 1 and 2:}
In the lexicon of adjectives, four different sources contributed adjectives.
Among these sources, the user experiment provided a total of 33 unique adjectives.
Hyundai uses adjectives for measuring the physical performance of a car door, eight of these were usable for our purpose.
Thirteen adjectives were collected from prior literature \cite{bezat2007car, yilmazer2017understanding}.
After analyzing the above three sources, the authors included 14 more adjectives based on their experience and knowledge of working in this domain. 
They felt these could be useful additions to the lexicon of adjectives.
Combining all these sources, the lexicon contained a total of 68 adjectives, which are provided in Table~\ref{tab:lexiconofadjectives}.

The second experiment filtered out the most relevant adjectives for describing the perception of opening a door.
Each adjective was scored by the users, and these scores were averaged for all cars and users.
Figure \ref{fig:relevantadjectives} shows the relevance of each adjective. 
It was empirically decided to choose the adjectives that were selected by at least 20$\%$ of the users.
A total of 25 out of the 68 adjectives were selected based on this criterion.
These were further used in experiment 3.

\begin{table}
\centering
\caption{The ten adjective pairs used for the adjective rating experiment. Six of the adjectives were combined with adjectives with similar perceptual connotations.}
\label{tab:adjectivepairs}
\begin{tabular}{|cll|}
\hline
\multicolumn{3}{|c|}{Antonymous Adjective Pairs}                                               \\ \hline
\multicolumn{1}{|c|}{1}  & \multicolumn{1}{l|}{Jerky}               & Easy to operate/Easy     \\ \hline
\multicolumn{1}{|c|}{2}  & \multicolumn{1}{l|}{Smooth}              & Rough                    \\ \hline
\multicolumn{1}{|c|}{3}  & \multicolumn{1}{l|}{Frictional/Forceful} & Frictionless/Comfortable \\ \hline
\multicolumn{1}{|c|}{4}  & \multicolumn{1}{l|}{Hard-to-pull}        & Easy-to-open             \\ \hline
\multicolumn{1}{|c|}{5}  & \multicolumn{1}{l|}{Balanced}            & Unstable                 \\ \hline
\multicolumn{1}{|c|}{6}  & \multicolumn{1}{l|}{Soft}                & Hard/Rigid               \\ \hline
\multicolumn{1}{|c|}{7}  & \multicolumn{1}{l|}{Cheap/old}           & Classy                   \\ \hline
\multicolumn{1}{|c|}{8}  & \multicolumn{1}{l|}{Damped}              & Recoiling                \\ \hline
\multicolumn{1}{|c|}{9}  & \multicolumn{1}{l|}{Discordant}          & Consistent               \\ \hline
\multicolumn{1}{|c|}{10} & \multicolumn{1}{l|}{Heavy/Stiff}         & Light                    \\ \hline
\end{tabular}
\end{table}

\begin{figure}[t]
        \centering
        \includegraphics[width=1\columnwidth]{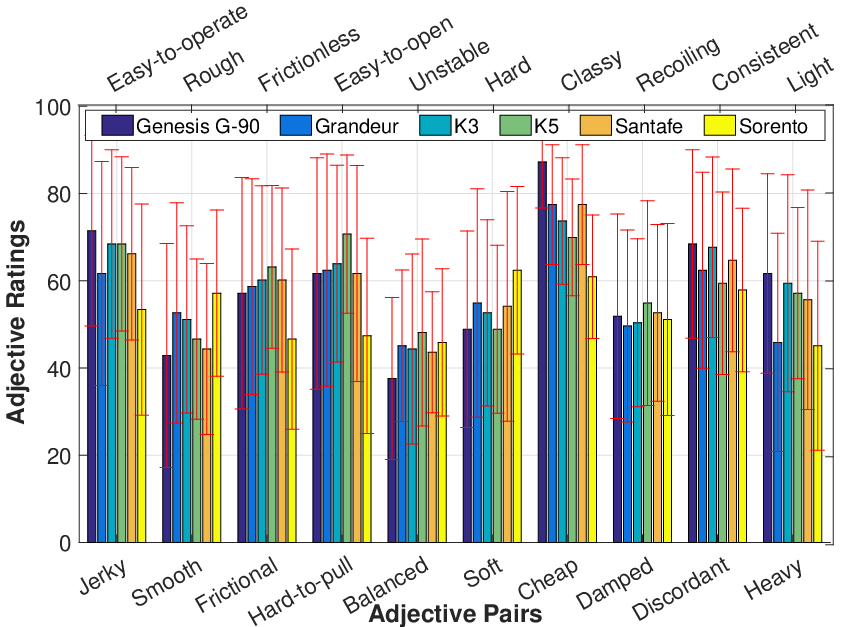}
        \caption{Averaged adjective rating for ten adjective pairs from experiment 3. The error bars show the standard deviation for each bar.}
        \label{fig:adjectiveratings}
\end{figure}

\begin{figure*}
        \centering
        \includegraphics[width=1.95\columnwidth, height = 1.2\columnwidth]
        {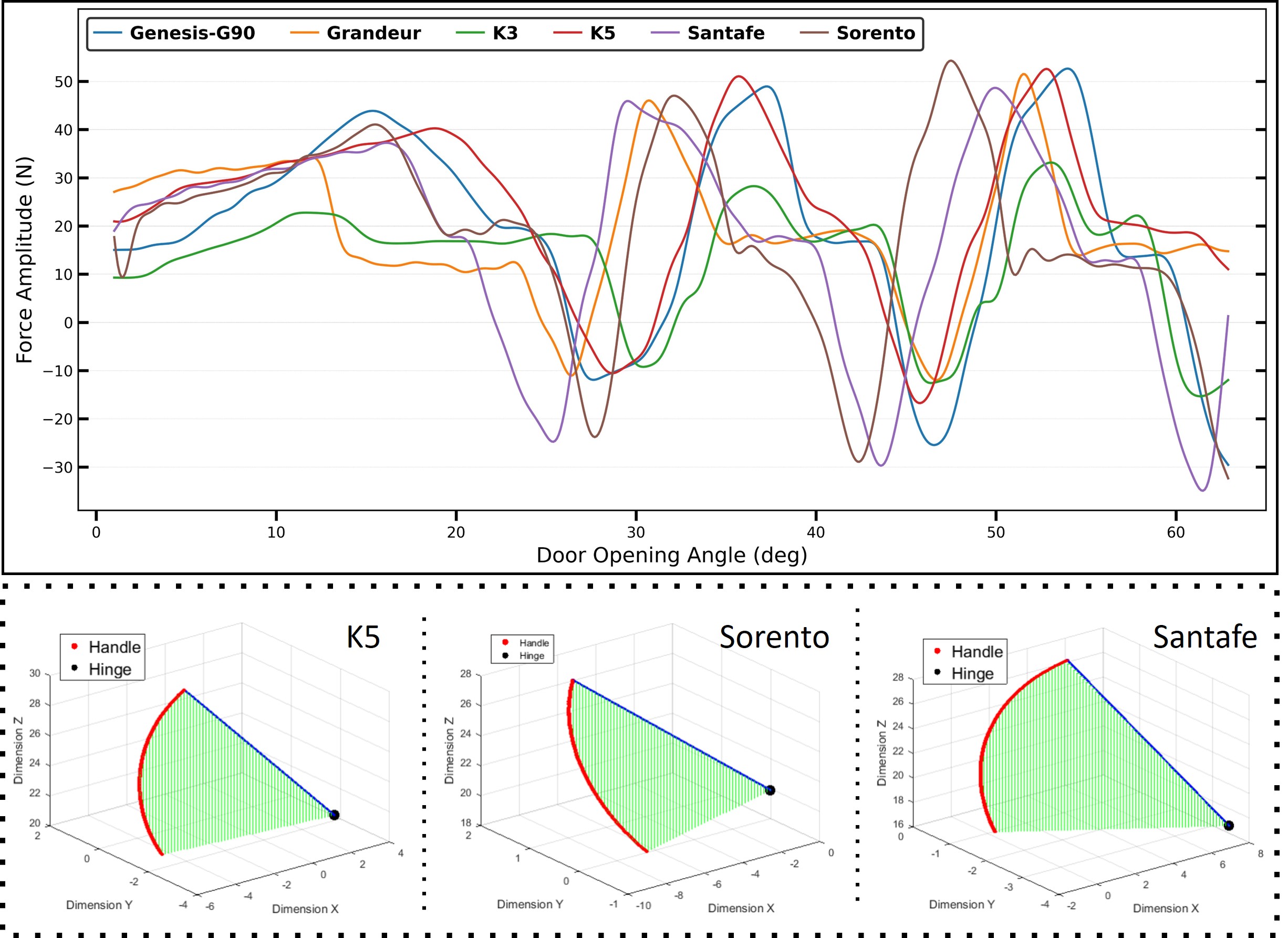}
        \caption{Angle-normalized force profiles of the six cars used in this study (Top). The position tracking of the door opening is provided for K5, Sorento, and Santafe for reference (Bottom).}
        \label{fig:carprofiles}
\end{figure*}

\subsection{Experiment 3: Adjective Rating}
\label{subsec:experiment3}
The 25 adjectives selected after the second experiment were divided into pairs of antonymous attributes to represent the opposite ends of the same scale.
Six adjectives were similar in meaning and paired at the same end of the scale.
Heavy received the highest score from users in experiment 2, however, there was no straightforward antonymous pairing available from within the top 25 adjectives. 
Light was selected from outside the top 25 to pair with heavy.
A total of ten pairs were formed at the end of this exercise, as shown in Table \ref{tab:adjectivepairs}.

In this experiment, users were provided with a list of the selected ten adjective pairs located at the opposite ends of a seven-point Likert scale.
The task was to rate the perception of opening a car door using these pairs.
The same procedure of opening the car door was followed as in the earlier experiments.
Each user rated all six cars in a randomized sequence.

% \textbf{Results of Experiment 3:}

\vspace{1em}
\noindent \textbf{Results of Experiment 3:}
The data from experiment 3 were in the form of adjective ratings for six cars rated against ten adjective pairs.
The data were averaged for all users and normalized onto a scale of zero to 100.
Adjective rating data for all cars and adjective pairs are shown in Fig. \ref{fig:adjectiveratings}.

\section{Force Profile of Opening a Car Door}
\label{sec:forceprofile}
In the psychophysical experiments users opened a car door and provided perceptual ratings. 
The perceptual characteristics exhibited by an opening car door are highly dependent on the physical aspects of the door.
Therefore, a physical signal that can describe the act of opening a door should be considered significant.
The force profile can be considered an important physical aspect of opening a door.
It refers to the amount of force required to open (and close) the door at different points in its range of motion.
It takes into account several factors that contribute to the perceptual characteristics of a car door.
It can be considered as the combined effect of the weight of the door, its aerodynamics, and the shape of the hinge that keeps it attached to the main frame.
Therefore, it was decided to use the force profile for predicting the perceptual characteristics of opening a door.
In the current study, force profiles of the cars provided in \ref{subsec:participantsanddataset} were recorded.

\subsection{Data Collection Setup}
\label{subsec:datacollectionsetup}
To record the force profile of the car door, we used an ATI force sensor and an Optitrack Trio120 optical sensor. 
The ATI force sensor was attached to the door handle, and Optitrack markers were placed just beside the handle so that they were visible to the cameras at all times.
A one-time position tracking of the door hinge was carried out for every car.
This was done to establish a reference point for measuring the opening angle.
A user opened the door with their left hand.
The users were instructed to make a conscious effort to maintain a constant velocity and  avoid jerks.
The force sensor recorded the force required to open the door at different points in its range of motion. 
The Optitrack Trio 120 was used to track the movement of the door and the markers to provide a visual representation of the door's range of motion.
The setup is presented in Fig. \ref{fig:framework}.
The data from both sensors were synchronized based on timestamps.
The force sensor recorded data at 1 kHz while Optitrack provided position data at an update rate of 80 Hz. 
The position data were upsampled to match the force sensor update rate. 
A total of ten force profiles were recorded for each car.

\subsection{1D Force Profiles}
\label{subsec:1dforceprofiles}

The data collected from different cars was inconsistent because it was collected by human users. The maximum opening angles of the cars were also variable. To make the data more comparable and accurate, it was important to normalize it and make it uniform across all cars.

The maximum opening angle for all the cars was capped at 63$^\circ$, as most cars had a maximum opening angle below this limit.
For cars with smaller maximum angles, data were zero-padded at the end.
Since data were collected by human users, the opening velocity was variable.
This was normalized by combining the position tracking data and force data.
The force data were divided into subsets corresponding to a range of 1$^\circ$ of the angle. 
The subset of force data for each degree was then downsampled and truncated to 10 data points. 
This was done to make the data uniform across all cars and smooth out outliers.
A total of ten force data points were selected for each degree of opening the car angle, resulting in a total of 630 data points for each car profile. 
A total of 10 recordings were carried out per car, to provide multiple training instances of the same data for the deep learning model.
Force profiles of all six car doors and position tracking of three car doors are provided in Fig.~\ref{fig:carprofiles}.

\section{CNN-LSTM Network }
\label{sec:cnnlstmnetwork}

\begin{figure*}[t]
\centering
        \includegraphics[width=2.0\columnwidth]{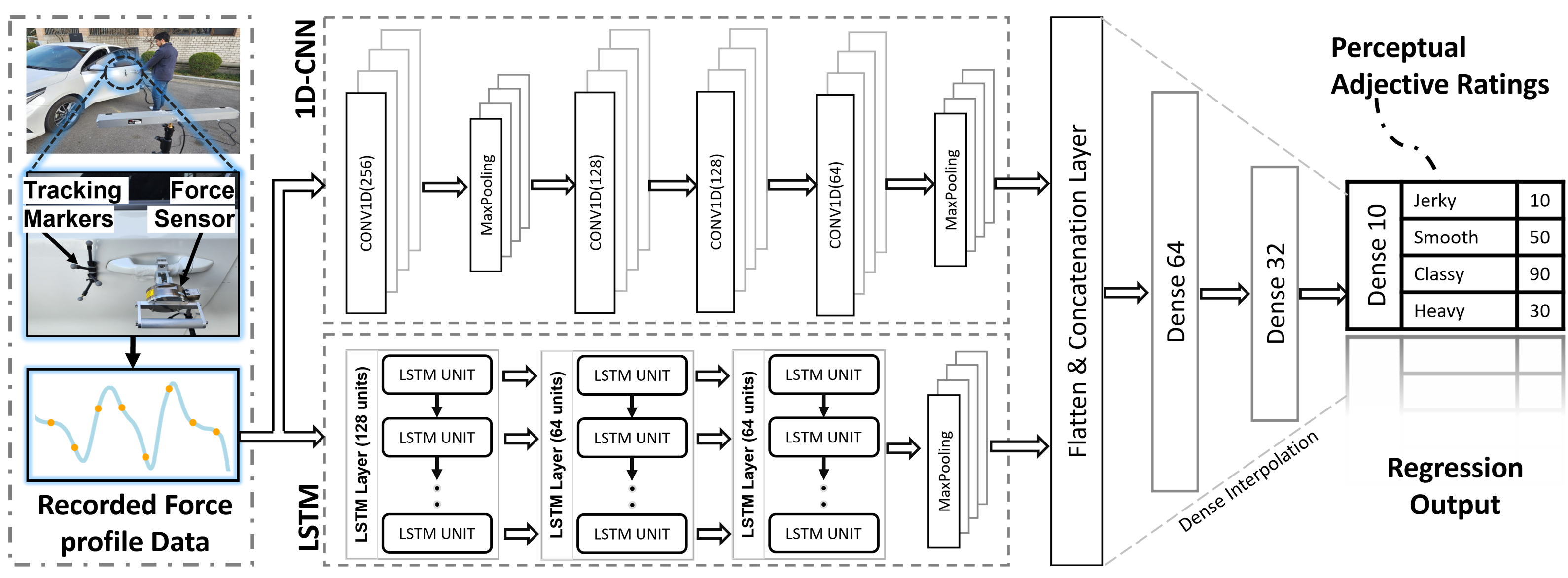}
        % \caption{A block diagram of the overall framework. The top row details the steps required to establish the HAS which is the first main contribution of this work. The next two rows show the training and testing methodology of the 1D-CNN.}
        \caption{The architecture of the proposed CNN-LSTM network. This model takes the force profile of the door as input and predicts the human-perceptual ratings associated with it.}
        \label{fig:model}
\end{figure*}

Statistical approaches, such as AR (Auto-regressive), MA (Moving Average), ARMA (Auto-regressive Integrated Moving Average), and their other variants are widely used to process time series data, but these methods do not always give the best results.  
The reason is that these approaches do not take into account long-term temporal dependencies \cite{awan2023predicting}. 
While deep learning approaches, such as recurrent neural networks (RNN), can effectively process time-series data. However, even RNNs have their own set of challenges, especially when it comes to dealing with long input sequences. 
In such a case, RNN can face a vanishing gradient problem during back-propagation. 
This problem is well addressed by the Long Short-Term Memory (LSTM) network and exhibited notable performance in detecting long-short-term temporal dependencies \cite{hochreiter1997long},\cite{yang2020cnn}. 
Likewise, Convolutional Neural Network (CNN) showed good prediction accuracy in numerous applications related to image and speech processing such as image segmentation \cite{kayalibay2017cnn} and speech-emotion recognition \cite{qayyum2019convolutional} respectively by extracting spatial information.

Recent work has applied deep learning to haptic signal processing for tasks such as surface texture classification \cite{zheng2016deep}, high-frequency vibration synthesis \cite{tao2024cross}, haptic attribute estimation \cite{awan2023predicting}, and perceptual similarity modeling \cite{priyadarshini2019perceptnet}. These advances highlight the potential of combining temporal and spatial modeling to better understand and interpret haptic signals.

Motivated by these insights, this study proposes a hybrid CNN-LSTM network that leverages the pattern extraction capabilities of convolutional layers and the sequence modeling strength of LSTMs. The objective is to predict perceptual attributes of car doors based on the force signals generated by door hinge dynamics, as described in Section~\ref{sec:forceprofile}. In this formulation, the one-dimensional force profiles are treated as time-series signals, allowing the model to learn both short-term variations and longer-range dependencies that may arise throughout the door-opening motion. To facilitate this, the architecture is designed as a two-stream network operating in parallel, where each stream can be considered a modular unit composed of a 1D-CNN followed by an LSTM. This structure enables the model to extract meaningful features from different aspects of the input and jointly contribute to the final prediction.
The following section describes the architecture of the proposed 1D-CNN and LSTM modules.

\subsection{1D Convolutional Neural Network}

Convolutional Neural Networks (CNNs) are widely used for analyzing structured signals due to their ability to extract hierarchical features through local receptive fields. In this work, a 1D-CNN module is employed to extract spatial features from the input force profile, which is treated as a 1D time-series signal. This module is particularly adept at identifying localized patterns in the profile, such as sharp force changes or periodicities, which may correspond to hinge geometry, resistance variations, or detents in the door mechanism. These features are critical for modeling the perceptual characteristics associated with door dynamics.

The architecture of the proposed 1D-CNN consists of four convolutional layers and two max pooling layers. The first convolutional layer takes the raw force profile of length 630 as input and applies 256 filters with a kernel size of 
1×3 and a stride of 1. A max pooling layer with a pool size of 2 follows to downsample the feature map and reduce computational complexity. The output is then passed through three additional convolutional layers with 128, 128, and 64 filters, respectively, each followed by non-linear activation. Another max pooling layer is inserted after the final convolutional layer to further reduce dimensionality and improve model generalizability. The final feature map from this CNN module is flattened and forwarded to the feature fusion layer for integration with the LSTM module.

\subsection{Long Short--Term Memory (LSTM)}

Long Short–Term Memory (LSTM) networks are a specialized class of recurrent neural networks (RNNs) designed to capture long-range temporal dependencies in sequential data. Unlike conventional RNNs, which suffer from vanishing or exploding gradients during training, LSTMs employ a memory cell and gating mechanisms to preserve information across extended time steps \cite{hochreiter1997long}. These properties make LSTMs well-suited for modeling time-varying physical signals such as force profiles, where the relationship between earlier and later values in the sequence is often meaningful.

In contrast to the 1D-CNN module, which is effective at capturing local spatial features and abrupt changes in the force signal, the LSTM is designed to learn global temporal dependencies that develop across the full door-opening trajectory. By combining these complementary capabilities, the overall model benefits from both local pattern detection and long-term sequence modeling.

The LSTM module in the proposed architecture consists of three stacked layers. The first LSTM layer contains 128 units and processes the input force profile directly. This is followed by two additional LSTM layers, each with 64 units. A max pooling operation with a pool size of 2 is applied after the final layer to reduce the temporal resolution and to regularize the feature representation. The input to the LSTM module is a one-dimensional sequence of \( n = 630 \) samples, where each element corresponds to the force required to open the car door at a specific angular displacement.

The internal operations of an LSTM unit at time step \( t \) are defined by the following equations:

\begin{align}
    i_t  & = \sigma(W_i (x_t + h_{t-1}) + b_i) \\
    f_t  & = \sigma(W_f (x_t + h_{t-1}) + b_f) \\
    o_t  & = \sigma(W_O (x_t + h_{t-1}) + b_O) \\
    C_t  & = f_t \odot C_{t-1} + i_t \odot \tanh(W_c (x_t + h_{t-1})) \\
    h_t  & = o_t \odot \tanh(C_t)
\end{align}

Here, \( i_t \), \( f_t \), and \( o_t \) denote the input, forget, and output gates, respectively; \( C_t \) represents the cell state and \( h_t \) is the hidden state at time \( t \). The symbol \( \odot \) denotes element-wise multiplication, and \( \sigma \) refers to the sigmoid activation function. The matrices \( W \) and vectors \( b \) are trainable parameters of the network.

The resulting features from the final LSTM layer, after pooling, is flattened and is then forwarded for integration with the spatial features extracted by the CNN module.

\subsection{Model Training Method}

To integrate the spatial and temporal features extracted from the 1D-CNN and LSTM branches, their outputs are flattened and concatenated into a single feature vector. This combined representation is passed through two dense layers with 64 and 32 units, followed by a final dense layer of size 10 as illustrated in Fig.~\ref{fig:model}.

The ReLU activation function is employed throughout the network to introduce non-linearity, except in the final output layer, where a linear activation is used to support continuous value prediction. The model is trained using the Root Mean Square Error (RMSE) as the loss function, optimized via the Adam optimizer, with a training duration of 100 epochs.
The learning rate, being one of the most sensitive hyperparameters \cite{yang2020cnn}, is determined through extensive experimentation. The values 0.05, 0.01, and 0.001 were evaluated through cross-validation to identify the most effective learning rate. Among these, the learning rate of 0.001 consistently provided stable training and the lowest validation error and was therefore selected as the optimal choice for this model. Among them, the learning rate of 0.001 demonstrated optimal convergence and was selected for the final model configuration.

\begin{figure*}[t]
  % include first image
\begin{subfigure}{.33\textwidth}
  \centering
  \includegraphics[width=1\linewidth]{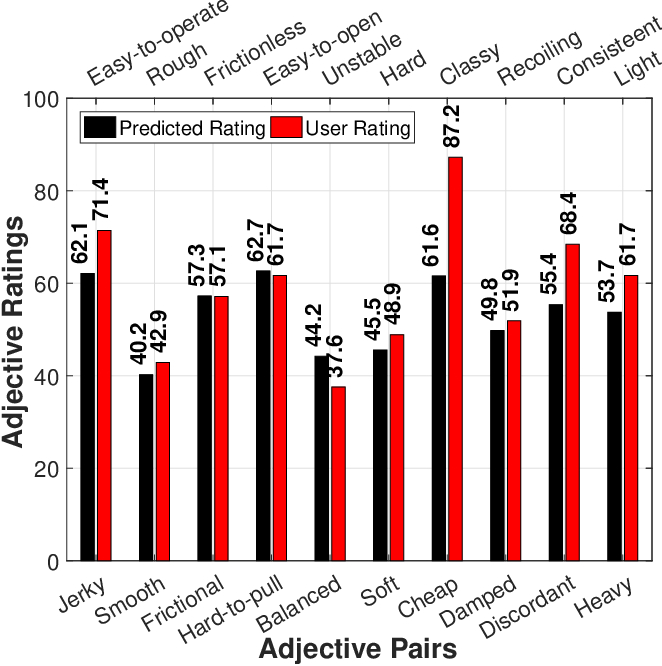}  
  \caption{Hyundai Genesis G90}
  \label{fig:sub-first}
\end{subfigure}
  % include second image
\begin{subfigure}{.33\textwidth}
  \centering
  \includegraphics[width=1\linewidth]{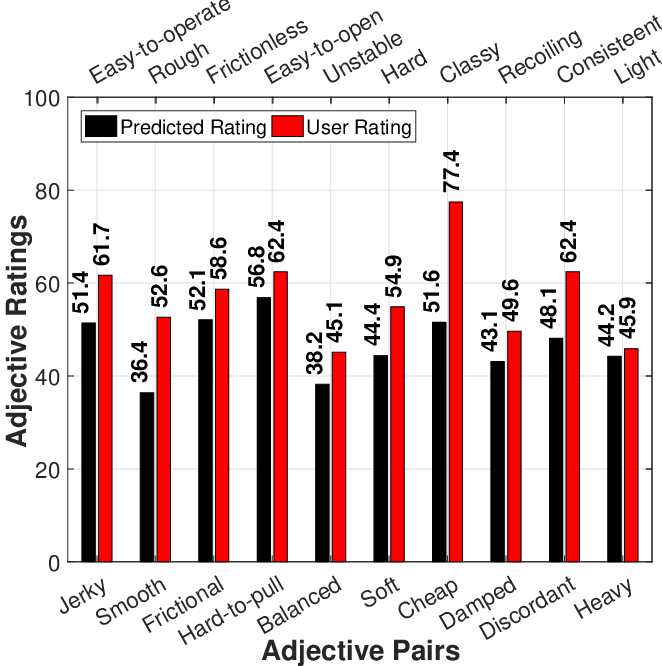}  
  \caption{Hyundai Grandeur}
  \label{fig:sub-second}
\end{subfigure}
 % include third image
\begin{subfigure}{.33\textwidth}
  \centering
  \includegraphics[width=1\linewidth]{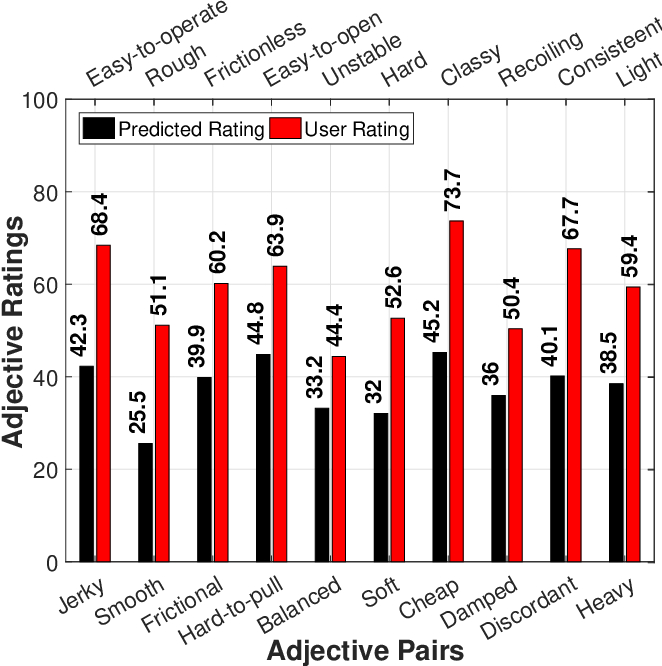}  
  \caption{Kia K3}
  \label{fig:sub-third}
\end{subfigure}
 % include fourth image
\begin{subfigure}{.33\textwidth}
  \centering
  \includegraphics[width=1\linewidth]{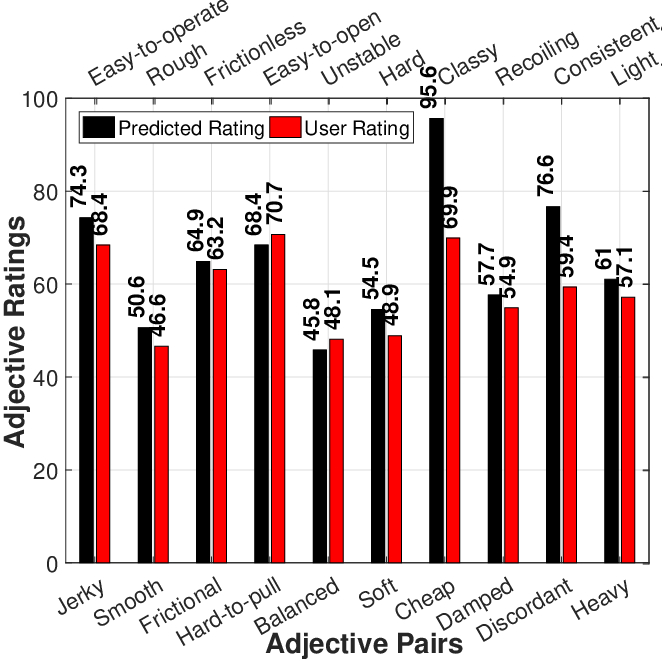}  
  \caption{Kia k5}
  \label{fig:sub-fourth}
\end{subfigure}
 % include fifth image
\begin{subfigure}{.33\textwidth}
  \centering
  \includegraphics[width=1\linewidth]{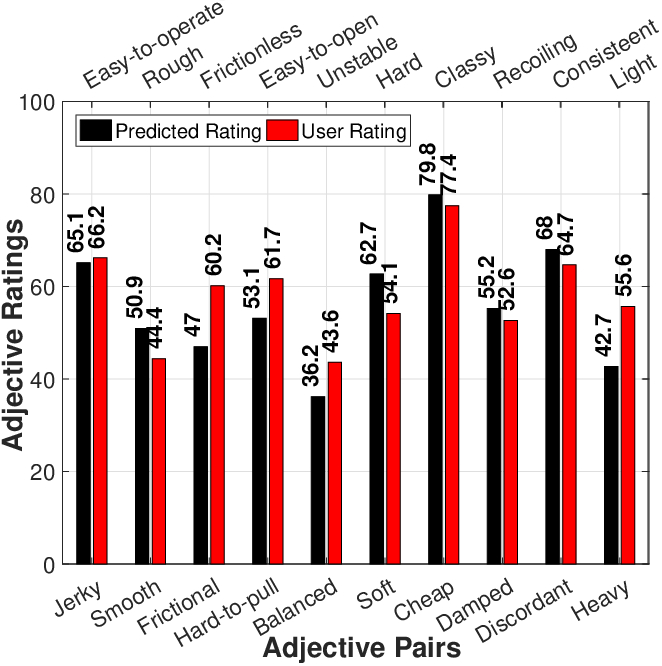}  
  \caption{Hyundai Santafe}
  \label{fig:sub-fifth}
\end{subfigure}
 % include sixth image
\begin{subfigure}{.33\textwidth}
  \centering
  \includegraphics[width=1\linewidth]{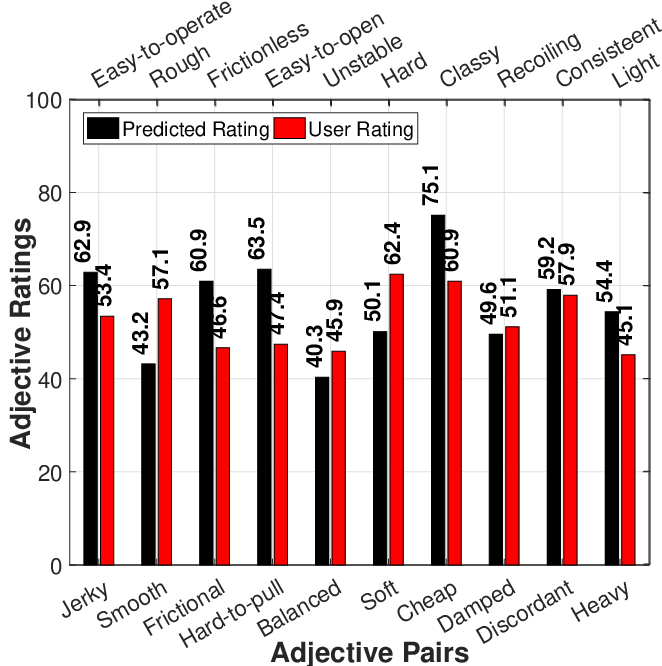}  
  \caption{Kia Sorento}
  \label{fig:sub-sixth}
\end{subfigure}
\caption{Leave-one-out Cross-Validation results for each of the six cars used in this study. The predicted and human-rated values are presented for ten adjective pairs.}
\label{fig:results}
\end{figure*}

\section{Evaluation}
\label{sec:evaluation}

The purpose of the system under consideration is to precisely predict the haptic perception of opening a car door through the analysis of its force profile. 

In order to gauge the model's ability to predict door-opening attributes for unseen force profiles, a numerical evaluation is conducted using Leave-One-Out Cross-Validation (LOOCV). 
This evaluation gauges the system's ability to predict haptic attribute values for force profiles it has not encountered before, measuring its predicting proficiency.

\begin{table*}
\caption{Mean absolute error (MAE) For all Cars and Adjective Pairs using the LOOCV method.}
\label{tab:mae}
\begin{tabular}{|c|cccccccccc|c|}
\hline
               & \begin{tabular}[c]{@{}c@{}}Jerky\\ Easy-to-operate\end{tabular} & \begin{tabular}[c]{@{}c@{}}Smooth\\ Rough\end{tabular} & \begin{tabular}[c]{@{}c@{}}Frictional\\ Frictionless\end{tabular} & \begin{tabular}[c]{@{}c@{}}Hard-to-pull\\ Easy-to-open\end{tabular} & \begin{tabular}[c]{@{}c@{}}Balanced\\ Unstable\end{tabular} & \begin{tabular}[c]{@{}c@{}}Soft\\ Hard\end{tabular} & \begin{tabular}[c]{@{}c@{}}Cheap\\ Classy\end{tabular} & \begin{tabular}[c]{@{}c@{}}Damped\\ Recoiling\end{tabular} & \begin{tabular}[c]{@{}c@{}}Discordant\\ Consistent\end{tabular} & \begin{tabular}[c]{@{}c@{}}Heavy\\ Light\end{tabular} & \textbf{\begin{tabular}[c]{@{}c@{}}MAE\\ \%\end{tabular}} \\ \hline
Genesis        & 9.30                                                            & 2.63                                                    & 0.12                                                                 & 1.01                                                                   & 6.62                                                           & 3.32                                                   & 25.63                                                     & 2.11                                                          & 13.01                                                              & 7.91                                                    & \textbf{7.17}                                              \\
Grandeur       & 10.25                                                           & 16.25                                                   & 6.55                                                                 & 5.56                                                                   & 6.92                                                           & 1.51                                                   & 25.86                                                     & 6.51                                                          & 14.30                                                              & 1.63                                                    & \textbf{10.43}                                             \\
K3             & 26.15                                                           & 25.58                                                   & 20.28                                                                & 19.09                                                                  & 11.17                                                          & 20.60                                                  & 28.43                                                     & 14.41                                                         & 27.52                                                              & 20.90                                                   & \textbf{21.41}                                             \\
K5             & 5.88                                                            & 3.98                                                    & 1.70                                                                 & 2.24                                                                   & 2.30                                                           & 5.63                                                   & 25.72                                                     & 2.78                                                          & 17.24                                                              & 3.89                                                    & \textbf{7.14}                                              \\
Santafe        & 1.05                                                            & 6.52                                                    & 13.19                                                                & 8.53                                                                   & 7.45                                                           & 8.56                                                   & 2.36                                                      & 2.56                                                          & 3.31                                                               & 12.97                                                   & \textbf{6.65}                                              \\
Sorento        & 9.46                                                            & 13.98                                                   & 14.32                                                                & 16.12                                                                  & 5.57                                                           & 12.30                                                  & 14.20                                                     & 1.57                                                          & 1.26                                                               & 9.23                                                    & \textbf{9.80}                                              \\ \hline
\textbf{MAE \%} & \textbf{10.35}                                                 & \textbf{11.49}                                          & \textbf{9.36}                                                        & \textbf{8.76}                                                          & \textbf{6.67}                                                  & \textbf{10.15}                                         & \textbf{20.36}                                            & \textbf{4.99}                                                 & \textbf{12.78}                                                     & \textbf{9.42}                                            & \textbf{10.43}                                             \\ \hline
\end{tabular}
\end{table*}

\subsection{Leave-One-Out Cross Validation}

Cross-validation is a powerful technique for assessing a model's predictive performance on unseen data. 
It evaluates the model's ability to generalize its learning from the training data to new, unseen data. 
One form of cross-validation is k-fold cross-validation, where the data is divided into k subsets and a fixed number of subsets are used for training while the rest are used for testing. 
This process is repeated until all subsets have been used for testing. 
Leave-One-Out Cross-Validation (LOOCV), a specific type of k-fold cross-validation with k = 1, trains the model on all instances except for one, which is used as the test data. 
This method comprehensively evaluates the model, ensuring that every item in the dataset is used as a test case.
% \textcolor{blue}{This method comprehensively evaluates the model, ensuring that every item in the dataset is used as a test case.}
LOOCV can be considered as a computation-heavy evaluation method, however, it was selected for this study's in-depth evaluation of the proposed model, as the dataset used is not considered large in the machine learning field.

The dataset described in the Sec: \ref{subsec:participantsanddataset}, consisting of force profiles and user adjective ratings for six cars, was employed for LOOCV. 
According to LOOCV, the model was trained using five cars in the dataset, with the remaining one as the test set.
% \textcolor{blue}{According to LOOCV, the model was trained using five cars in the dataset, with the remaining one as the test set.}
However, after running initial tests, it was noted that the force profile of Kia K3 was significantly different from all other profiles, and the prediction accuracy was reduced if K3 was used for training.
Therefore, K3 was not used in training the model, instead, the model was trained using the data of four cars at a time, and the fifth one was used as the test set (with K3 being left out every time). 
K3 was also used as a test set, where the model was trained using only four other cars to keep the training data size consistent (Santafe was randomly chosen to be excluded).
This process was repeated until all cars had been used as test sets. 
The prediction results from LOOCV for the proposed model are illustrated in Fig. \ref{fig:results}.

% \begin{figure*}
% \centering
%         \includegraphics[width=2.0\columnwidth]{figs/Density_Error_plot.jpg}
%         \caption{Need to be written.}
%         \label{fig:Density_error}
% \end{figure*}

The Mean Absolute Error (MAE) was calculated for all the adjective pairs and all the cars to better understand the prediction results, as shown in Table \ref{tab:mae}. 
The MAE offers a more direct and intuitive summary of the prediction results.
Table \ref{tab:mae} shows the individual prediction accuracy for each car against each of the adjective pairs.
The MAE $\%$ column on the right shows the averaged prediction error for each car, while the MAE $\%$ column at the bottom shows the averaged prediction error for each adjective pair.
It can be seen that the average prediction accuracy for most of the cars and adjective pairs is around 10 $\%$ or below. 
The only outliers (high prediction MAE) are the averaged results for K3 (21.41 $\%$), and the adjective-pair of Cheap-Classy (20.36 $\%$).

\begin{figure}[t]
        \centering
        \includegraphics[width=1\columnwidth]{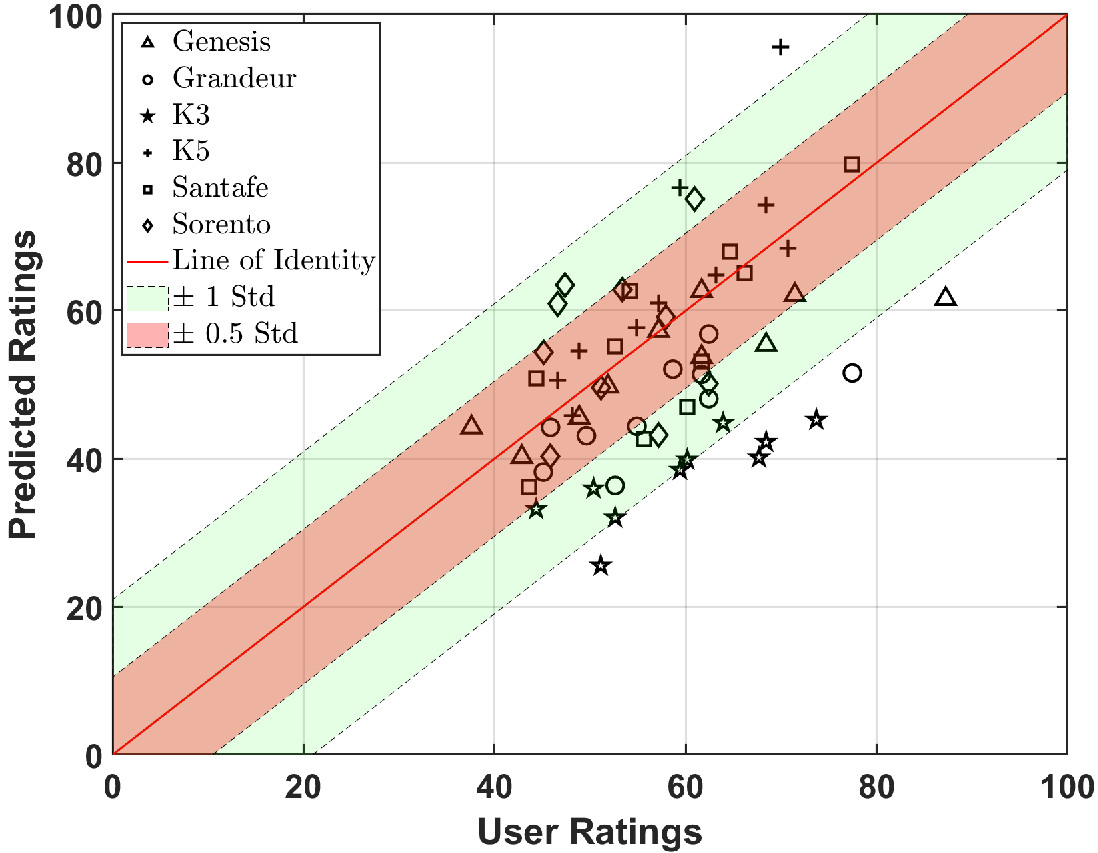}
        \caption{Analysis of the predicted ratings based on standard deviation in user ratings from the perspective of different cars in the dataset. The red line indicates a perfect prediction of the user rating by the algorithm. The red and green bands represent a half and the first standard deviation of the user ratings.}
        \label{fig:stdbandscars}
\end{figure}

\subsection{Error Analysis}
\label{sec:erroranalysis}
\begin{figure}[t]
        \centering
        \includegraphics[width=1\columnwidth]{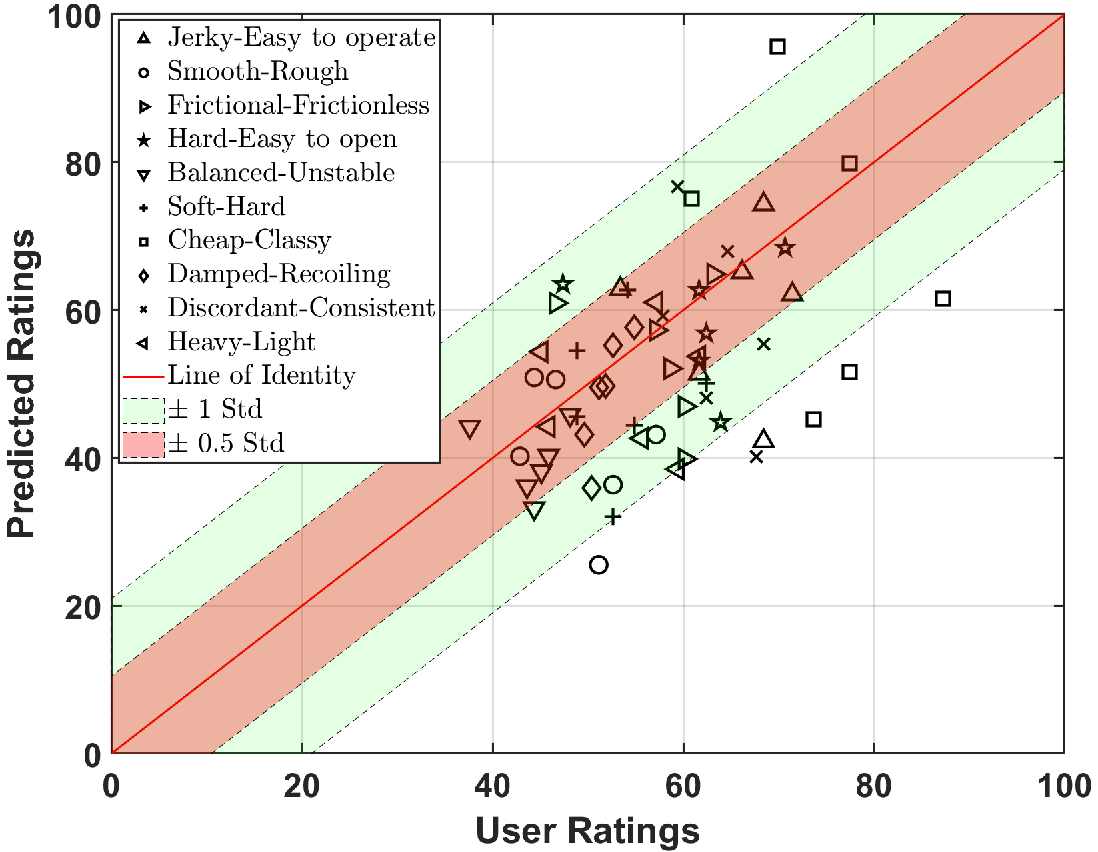}
        \caption{Analysis of the predicted ratings based on standard deviation in user ratings from the perspective of the adjective pairs. The red line indicates a perfect prediction of the user rating by the algorithm. The red and green bands represent a half and the first standard deviation of the user ratings.}
        \label{fig:stdbandsadj}
\end{figure}

Figures \ref{fig:stdbandscars} and \ref{fig:stdbandsadj} show an analysis of the predicted results in terms of the standard deviation of user ratings.
The x-axis represents the user ratings, and the y-axis represents the corresponding prediction by the algorithm.
In an ideal scenario all the data would be located on the diagonal line of identity (red line), shown in Figs. \ref{fig:stdbandscars} and \ref{fig:stdbandsadj}, where the predicted values and the user ratings would be the same.
However, in the current case values are scattered around this trend line due to prediction errors.
A point above the line of identity would signify that the algorithm under-predicted the user rating, while a point below the line of identity means that the predicted value was above the user rating.

Analysis of the user ratings from Sect. \ref{subsec:experiment3} shows that the user ratings contained variations across participants.
These variations are expected as haptic perception can vary from one person to another.
In order to account for these variations, the average standard deviation across the six cars and the adjective pairs was calculated for all participants.
The standard deviation for the cars was 22.0 for Genesis G90, 22.89 for Grandeur, 20.95 for K3, 19.30 for K5, 20.54 for Santafe, and 20.11 for Sorento.
Similarly, the average standard deviation for the adjective pairs was 22.17 for Jerky-Easy to operate, 21.54 for Smooth-Rough, 22.20 for Frictional-Frictionless, 23.48 for Hard-Easy to open,  18.31 for Balanced-Unstable, 22.46 for Soft-Hard, 13.34 for Cheap-Classy, 21.73 for Damped-Recoiling, 20.88 for Discordant-Consistent, and 23.57 for Heavy-Light.
The standard deviation averaged for all cars or all adjective pairs was 20.96.
The red and green bands in Figs. \ref{fig:stdbandscars} and \ref{fig:stdbandsadj} highlight the half and first standard deviation from the ideal prediction line.
It can be seen that a majority of the data points fall within the first standard deviation across both figures.
The consistent outliers in both cases are the data points for K3 among cars, and Cheap-Classy among adjective pairs, as expected.

\section{Discussion}
\label{sec:discussion}

We developed a CNN-LSTM model capable of predicting users' haptic perceptions of car doors based on their force profiles. 
The model achieved a prediction MAE of around 10\%, indicating its effectiveness in translating physical interaction data into subjective user evaluations. 
These findings suggest that our system can help designers and engineers assess the perceptual attributes of car doors in the early stages of development, reducing the reliance on physical prototypes.

\subsection{Translating Force Profiles into Cognitive Attributes}

In this study, we established a link between the \textit{Torque/Force Profile Space} and the \textit{Cognitive Attributes Space} for car doors. 
On one hand. the \textit{Torque/Force Profile Space} represents the physical interaction data generated when opening car doors, which we collected using force sensors and optical trackers. 
This data captures physical dynamics such as the weight, resistance, and smoothness of door operation.
On the other hand, we defined a \textit{Cognitive Attributes Space} based on user feedback about the physical attributes. 
Participants were asked to rate their haptic experiences with car doors using a set of antonymous adjective pairs to quantify their subjective evaluations. 
This established a \textit{Cognitive Attributes Space} tailored to the unique perceptual dimensions of car door operation.
Finally, we used a CNN-LSTM model to connect these two spaces. 
The model associated the physical dynamics of car doors with cognitive attributes, demonstrating that the force profiles contain sufficient information to accurately predict user perceptions.

\subsection{Interpreting Prediction Errors}

Most predictions showed an MAE of around 10\% or lower (Fig. \ref{fig:results}, Table \ref{tab:mae}). 
Although the Just Noticeable Difference (JND)\textemdash the smallest detectable difference between two stimuli \cite{gescheider2013psychophysics}\textemdash for haptic attributes is not explicitly available in the literature, previous work suggests that perceptual boundaries are not sharply defined \cite{hassan2017perceptual}.
% Although the Just Noticeable Difference (JND)\textcolor{blue}{\textemdash the smallest detectable difference between two stimuli \cite{gescheider2013psychophysics}\textemdash} for haptic attributes is not explicitly available in the literature, previous work suggests that perceptual boundaries are not sharply defined \cite{hassan2017perceptual}.
To estimate how much MAE is perceptually acceptable, we calculated the average standard deviation of participant ratings, which was 20.96. 
Since this deviation represents the natural variability in user perception, a prediction MAE of 10\% likely falls within the range of perceptual similarity and can be considered insignificant.

Some model predictions deviated significantly from the ground truth, likely due to non-linear or complex relationships between input features and adjective pairs. 
User bias, such as preference for a specific car or misunderstanding of adjective pairs, may have influenced these discrepancies.
Similarly, prejudice or admiration for a car model could skew the results.

\subsection{Limitations and Future Work}

In the current study, as an initial proof of concept, we used a small dataset of cars for training and evaluation.
The number of adjectives was extensive, and the use of six different cars provided a reasonable diversity; however, it could be a limiting factor.
Additionally, the use of real cars may have introduced visual bias in the perceptual ratings, as participants could not be fully isolated from the visual appearance of the cars, despite instructions to ignore it. 
A potential solution to both issues is the development of a door simulator, which could generate diverse force profiles and provide a controlled environment for consistent and unbiased data collection. 
This approach would allow for greater data diversity and more accurate modeling of the relationship between \textit{Torque/Force profiles} and the \textit{Cognitive Attributes}.

Future work in this domain could focus on expanding the dataset, including more adjectives, studying external factors that influence a user's perceptual ratings, refining the deep learning model or replacing it with a more suitable one, and considering more robust features from the force profile.

\section{Conclusion}
\label{sec:conclusion}

The current study presents a deep learning model for predicting the perceptual properties of opening a car door by analyzing force profiles.
The perceptual attributes were provided by human participants, whereas the force profiles were recorded by sensors attached to a car door.
The performance of the model was evaluated using LOOCV, and the results indicated a significant degree of accuracy in predicting perceptual attributes in most cases.
These findings highlight the potential applications of the model in the automotive industry for perceptual design evaluation of car doors.

% \section*{Acknowledgments}
% This should be a simple paragraph before the References to thank those individuals and institutions who have supported your work on this article.

\bibliographystyle{IEEEtran}  % Specifies the bibliography style
\bibliography{references}     % The filename of the .bib file, without the extension

\begin{IEEEbiography}[{\includegraphics[width=1in,height=1.25in,clip,keepaspectratio]{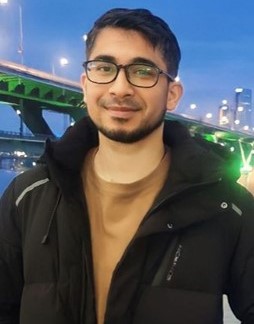}}]
{Mudassir Ibrahim Awan} 
received his B.E. in Electronics Engineering from the Karachi Institute of Economics and Technology (KIET), Karachi, Pakistan, in 2016. 
In 2018, he joined the Haptics and Virtual Reality Laboratory at Kyung Hee University, South Korea, where he is currently pursuing an integrated MS-PhD program in the Department of Computer Science and Engineering. His research interests include haptic modeling and rendering, psychophysics, drone haptics,  and car door perception.
\end{IEEEbiography}

\begin{IEEEbiography}
[{\includegraphics[width=1in,height=1.25in, clip,keepaspectratio]{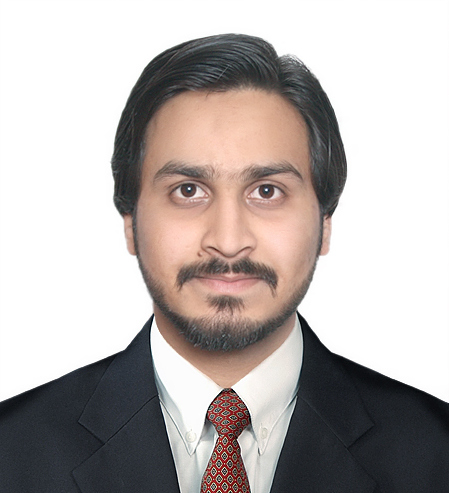}}]{Ahsan Raza} received a B.S. degree in computer engineering from the University of Engineering and Technology (UET), Taxila, Pakistan,  and the Ph.D. degrees in computer science and engineering
from Kyung Hee University, Republic of Korea in 2024.
His research interests include mid-air haptic feedback, haptic actuator design, and psychophysics.
\end{IEEEbiography}

\begin{IEEEbiography}[{\includegraphics[width=1in,height=1.25in,clip,keepaspectratio]{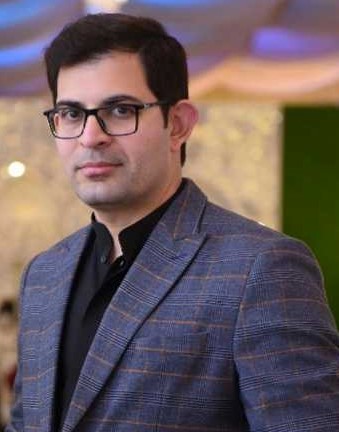}}]{Waseem Hassan} received the B.S. degree in electrical engineering from the National University of
Science and Technology (NUST), Pakistan, and the
M.S., and Ph.D. degrees in computer science and engineering
from Kyung Hee University, Republic of Korea in 2016 and 2022, respectively.
Currently, he is working as a Post-Doctoral researcher with the Department of Computer Science, University of Copenhagen, Denmark. 
His research interests include psychophysics, haptic content generation, and novel haptic interfaces.
\end{IEEEbiography}

\begin{IEEEbiography}
[{\includegraphics[width=1in,height=1.25in, clip,keepaspectratio]{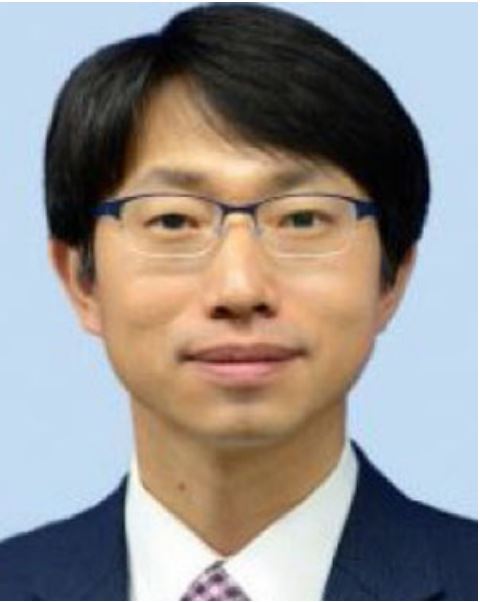}}]{Ki-Uk Kyung} 
(Member, IEEE) received the bachelor’s and Ph.D. degrees in mechanical engineering from the Korea Advanced Institute of Science and Technology (KAIST), Daejeon, South Korea, in 1999 and 2006, respectively.
Since 2018, he has been an Associate Professor of Mechanical Engineering and the Director of the Human–Robot Interaction Laboratory,
KAIST. In 2006, he joined the POSTPC Research Group, Electronics and Telecommunications Research Institute, Daejeon, South Korea, where he was the Director of the Smart UI/UX Device Research Section.
\end{IEEEbiography}

\begin{IEEEbiography}
[{\includegraphics[width=1in,height=1.25in, clip,keepaspectratio]{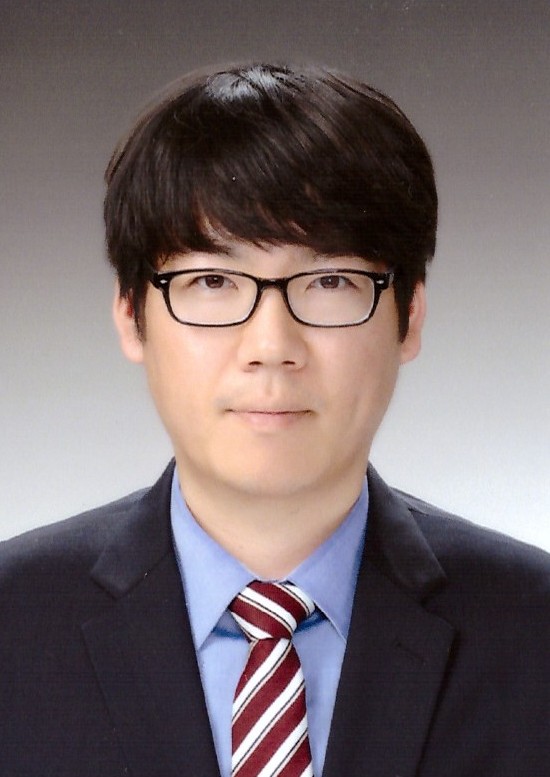}}]{Seokhee Jeon} (Member, IEEE) received the B.S.
and Ph.D. degrees in computer science and engineering from the Pohang University of Science and Technology (POSTECH) in 2003 and 2010, respectively.
He was a Post-Doctoral Research Associate with the
Computer Vision Laboratory, ETH Zurich. In 2012,
he joined the Department of Computer Engineering,
Kyung Hee University, as an Assistant Professor.
His research interests include haptic rendering in
an augmented reality environment, applications of
haptics technology to medical training, and usability
of augmented reality applications.
\end{IEEEbiography}

\EOD

\end{document}